\documentclass[aps,prl,twocolumn,titlepage,nofootinbib]{revtex4}
\usepackage{graphicx,physics}
\usepackage{color}
\usepackage{dcolumn}
\usepackage{latexsym}

\usepackage[normalem]{ulem}
\usepackage{hyperref,amssymb}
\usepackage{url}
\usepackage{color,soul}
\usepackage{graphicx}
\usepackage{verbatim}
\usepackage{multirow}
\usepackage{amsmath}
\newcommand{\beq}{\begin{eqnarray}}
\newcommand{\eeq}{\end{eqnarray}}
\usepackage{mathrsfs}
\usepackage{float,soul}
\usepackage[dvipsnames]{xcolor}
\usepackage{mathtools}
\usepackage{slashed}
\usepackage{physics}	
\usepackage{graphicx}   
\usepackage{epstopdf}
\usepackage{subfigure}  
\usepackage{hyperref}   
\usepackage{bbold}
\usepackage{wasysym}
\usepackage{feynmp}
\usepackage{hyperref}
\hypersetup{colorlinks,
}
\usepackage{bm}
\DeclareMathOperator{\sign}{sign}

\usepackage[latin1]{inputenc}

\begin{document}

\title{Designing atomic-scale resistive circuits in topological insulators\\ through vacancy-induced localized modes}
\author{Cunyuan Jiang$^{1,2}$}
\thanks{These authors contributed equally}
\author{Weicen Dong$^{1,2}$}
\thanks{These authors contributed equally}
\author{Matteo Baggioli$^{1,2}$}
\email{b.matteo@sjtu.edu.cn}
\address{$^1$School of Physics and Astronomy, Shanghai Jiao Tong University, Shanghai 200240, China}
\address{$^2$Wilczek Quantum Center, School of Physics and Astronomy, Shanghai Jiao Tong University, Shanghai 200240, China}

\begin{abstract}
We demonstrate that vacancies can induce topologically protected localized electronic excitations within the bulk of a topological insulator, and when sufficiently close, give rise to one-dimensional propagating chiral bulk modes. We show that the dynamics of these modes can be effectively described by a tight-binding Hamiltonian, with the hopping parameter determined by the overlap of electronic wave functions between adjacent vacancies, accurately predicting the low-energy spectrum. Building on this phenomenon, we propose that vacancies in topological materials can be utilized to design atomic-scale resistive circuits, and estimate the associated resistance as a function of the vacancy distribution's geometric properties. 
\end{abstract}
\maketitle

\color{blue}\textit{Introduction} \color{black} -- Over the past few decades, topological insulators (TIs), quantum materials distinguished by their insulating bulk energy gap and conducting, gapless surface states, have attracted significant attention from both theoretical, experimental and technological perspectives \cite{RevModPhys.82.3045,Rachel_2018,annurev:/content/journals/10.1146/annurev-conmatphys-062910-140432,Moore2010}. Due to their robust quantum transport properties and nontrivial topological characteristics \cite{PhysRevLett.95.146802,PhysRevLett.95.226801,PhysRevLett.96.106802,PhysRevLett.61.2015,Xu2014}, TIs have emerged as promising candidates for the next generation of electronic devices \cite{Breunig2022}.  

On the other hand, atomic-scale control of defects, deviations from perfect crystalline order, has become a cornerstone in modern materials design, with profound implications for mechanical performance, transport behavior, and electronic functionality. Understanding and manipulating such defects is therefore essential in both fundamental condensed matter physics and next-generation material technologies \cite{Xiong2021-oz,doi:10.1126/science.281.5379.939,Zhang2023,10.1246/cl.2010.1226}.

The interplay between electronic topology and crystalline defects gives rise to robust physical phenomena, mostly related to the emergence of topologically protected bound states localized around lattice imperfections. These states have been theoretically predicted and experimentally observed in a variety of systems, including topological insulators \cite{PhysRevB.89.161117,Lu2011,PhysRevX.11.011034,PhysRevB.91.075412,PhysRevB.84.035307,Lin2023,PhysRevLett.127.066401,PhysRevB.88.155127,Ran2009,PhysRevLett.124.243602,PhysRevLett.115.106601,PhysRevB.87.075126,PhysRevB.86.115433,PhysRevB.86.115327,PhysRevB.87.195122}. They arise because lattice defects, whether topological or not, can be regarded as boundaries within the material, attracting topological bound states similar to those formed at their surface via the bulk-edge correspondence \cite{Shen2017,PhysRevLett.124.243602,PhysRevLett.115.106601,PhysRevLett.125.240601,PhysRevB.88.155127,Ran2009}. Defects can be therefore manipulated to design wave guides for such protected topological modes. \textit{e.g.}, \cite{PhysRevLett.124.243602}.

Among all defects, point defects such as vacancies are generally easier to be created as they involve only the absence of an atom from a lattice site, rather than more complex rearrangements typical of other topological defects (dislocations, disclinations). Vacancies have been in fact used to manipulate properties of two-dimensional materials \cite{Mao2016}, hosting localized bulk states and strongly affecting the material conductivity \cite{PhysRevLett.107.116803,Chu_2012,PhysRevLett.129.196601}.

The effects of vacancies in topological insulators have been extensively studied in the literature. It has been demonstrated that the topological insulating phase remains robust in the presence of such lattice imperfections \cite{PhysRevB.101.125114}. Moreover, vacancy-induced bound states have been shown to emerge \cite{PhysRevB.84.035307}, along with the formation of higher-order topological corner states \cite{Tu_2023}. Here, we revisit this program in light of the promising opportunities offered by topological insulator-based devices \cite{Breunig2022} and their potential technological applications.

Specifically, we consider configurations of vacancies in a two-dimensional honeycomb lattice in the Haldane model of topological insulators \cite{PhysRevLett.61.2015}. We show that, when vacancies are sufficiently close, they give rise to propagating, chiral, and topologically protected excitations within the bulk of the topological insulator. We refer to these excitations as `\textit{vacanons}'. They originate from vacancy-induced electronic bound states that acquire mobility through hopping between neighboring vacancies. The dynamics of vacanons are governed by the topological nature of the host material via a \textit{bulk-defect correspondence}, with their chirality set by the Chern number. Moreover, their behavior can be accurately captured by an effective tight-binding Hamiltonian.

Leveraging on their dynamics, we propose the design of atomic-scale resistive circuits by engineering the inter-vacancy spacing, which directly controls the hopping amplitude of vacanons. This approach opens a novel pathway for utilizing topological insulators in atomic-scale electronic devices, offering distinct advantages over conventional semiconductor technologies in terms of robustness, scalability, and functional tunability.

\begin{figure}[ht!]
    \centering
    \includegraphics[width=\linewidth]{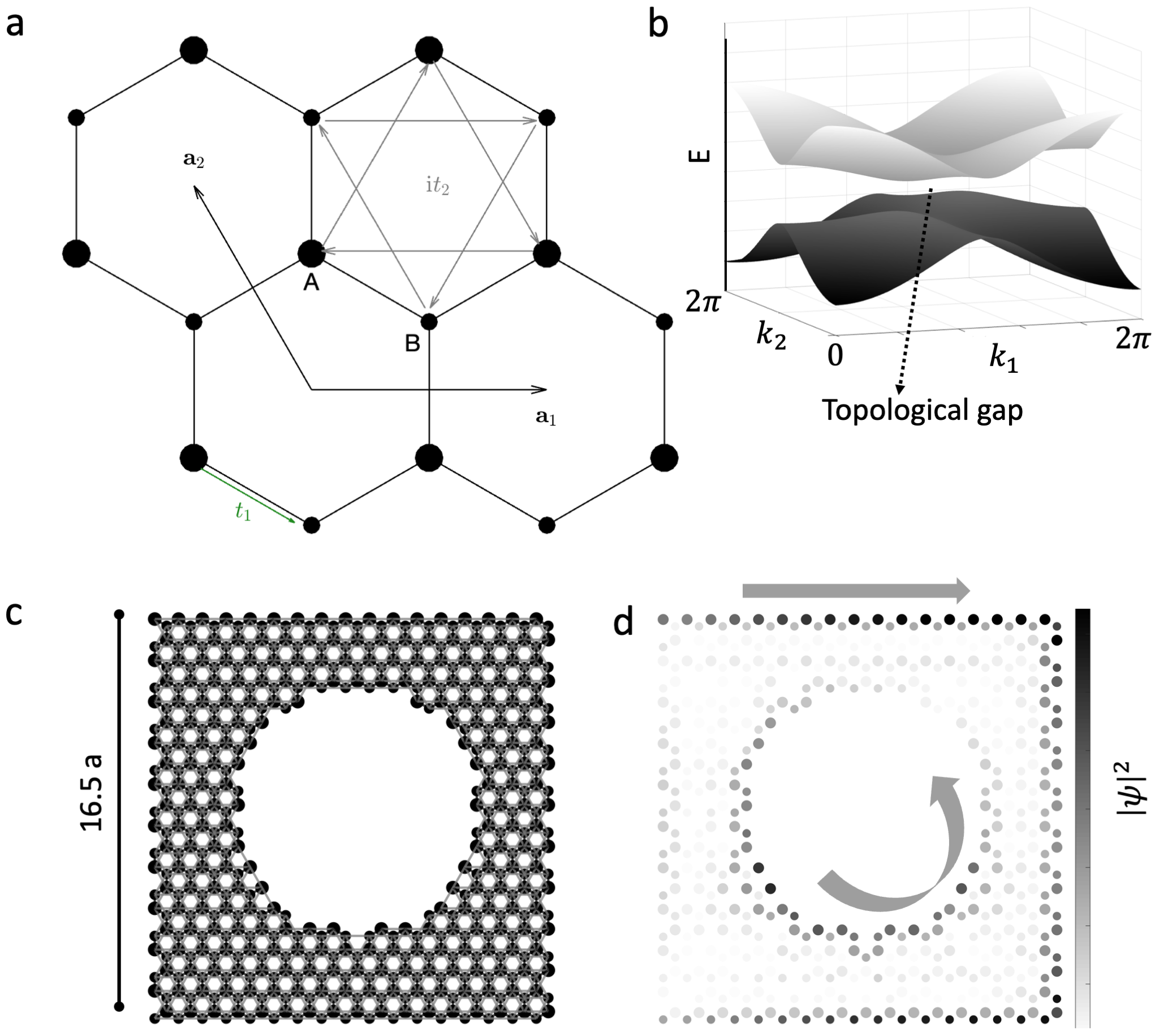}
    \caption{\textbf{(a)} 2D Honeycomb lattice with nearest-neighbor and next-nearest-neighbor hopping parameters indicated with green and gray arrows. The unit cell is defined by the unit vectors $\mathbf{a}_1=$a$(1,0)$ and $\mathbf{a}_2=$a$\left(-\frac{1}{2},\frac{\sqrt{3}}{2}\right)$. \textbf{(b)} Energy spectrum of the Haldane model for $t_1=1$ and $t_2=0.1$. The wave vector is expressed as $\textbf{k}=k_1 \textbf{b}_1 +k_2 \textbf{b}_2$ with $k_{1,2}$ the corresponding wave numbers and $\textbf{a}_i\cdot \textbf{b}_j=\delta_{i,j}$. \textbf{(c)} Real space lattice structure with artificial large cavity at the center of the sample. \textbf{(d)} Chiral states localized at the boundary of the sample and the edge of the hole, calculated using the structure shown in panel (c) with open boundary conditions. We set $t_1=1$ and $t_2=0.1$. The gray arrows indicate the direction of propagation while the background color represents the amplitude of the electronic wave-function $|\psi|^2$.}
    \label{fig1}
\end{figure}
\color{blue}\textit{Haldane model for topological insulators} \color{black} -- We consider the Haldane model on a honeycomb 2D lattice with real nearest-neighbor (NN) hopping, $t_1$, and imaginary next-nearest-neighbor (NNN) hopping, $\mathrm{i}t_2$~\cite{PhysRevLett.61.2015}. The tight-binding Hamiltonian in real space is given by:
\begin{equation}
    H = \sum_{\langle i,j\rangle} t_1  c_{i}^{\dagger} c_{j}+ \sum_{\langle\langle i,j\rangle\rangle} (\mathrm{i} t_2 c_{i}^{\dagger}c_{j}+\text{h.c.}),\label{simulation}
\end{equation}
where $c_{i}^{\dagger}$ and $c_{i}$ are respectively the creation/annihilation operators on the $i$-th lattice site. The NNN hopping parameter breaks time reversal symmetry and plays a fundamental role for the emergence of non-trivial topology within this model. The lattice structure and the relevant physical parameters are presented in Fig.~\ref{fig1}(a).

The Bloch Hamiltonian of the Haldane model in wave-vector ($\mathbf{k}$) space reads
 \begin{equation}
 \begin{aligned}
    H_\mathbf{k}&=t_1  \sum_j\left[\sigma_x \cos \left(\mathbf{k} \cdot  \bm{\delta}_j\right)-\sigma_y \sin \left(\mathbf{k} \cdot   \bm{\delta}_j\right)\right]
        \\&- t_2  \sigma_z \sum_j 2 \sin(\mathbf{k}\cdot  \bm{\delta}_j^\prime),
 \end{aligned}
    \end{equation}
where $\bm{\delta}_1=\frac{1}{3} \mathbf{a}_1-\frac{1}{3} \mathbf{a}_2$, $\bm{\delta}_2= \bm{\delta}_1+ \mathbf{a}_2$, and $\bm{\delta}_3= \bm{\delta}_1- \mathbf{a}_1$ are the vectors describing the NN hopping terms and $\bm{\delta}_1^\prime= \mathbf{a}_1$, $\bm{\delta}_2^\prime= -\mathbf{a}_1-\mathbf{a}_2$, and $\bm{\delta}_3^\prime= \mathbf{a}_2$ are the vectors representing the NNN hopping terms. When $t_2\neq0$, the energy bands present an insulating energy gap at the Fermi energy, as shown in panel (b) of Fig.~\ref{fig1}.

Most importantly, for $t_2 \neq 0$, the system is a topological insulator described by a finite Chern number, $C=-\sign(t_2)$. This is accompanied by the emergence of topologically protected edge states whose chirality is determined by the sign of $t_2$ ~\cite{PhysRevB.90.024412,PhysRevLett.71.3697}. For more details on this model, we refer to Haldane's lectures \cite{lec}.

\color{blue}\textit{Defect-induced localized topological bulk states} \color{black} -- The presence of structural defects, distorting local translational symmetry, can strongly alter the properties of TIs, resulting for example in finite bulk conductivity. Most importantly, impurities and defects can be regarded as system boundaries, allowing for the emergence of topological protected bulk modes \textit{akin} to boundary edge states \cite{Shen2017}.

In order to verify this phenomenon, we distort the 2D honeycomb lattice by placing a large cavity at its center, see Fig. \ref{fig1}(c). We then consider the time evolution of an initial state $| \phi(0) \rangle$, 
\begin{equation}
    | \phi(t) \rangle = e^{- i H  t}
     | \phi(0) \rangle,
\end{equation}
where $| \phi(0) \rangle$ is constructed from an arbitrary linear superposition of the real space Hamiltonian's eigenstates that lie within the topological gap. The resulting time evolution with open boundary conditions is presented in panel (d) of Fig. \ref{fig1} and it exhibits the presence of two localized chiral states (see attached videos). The first one is the common chiral edge mode emerging in the original Haldane model via the celebrated bulk-edge correspondence. On the other hand, a second mode appears along the boundary of the cavity made in the honeycomb structure (panel (c) in Fig. \ref{fig1}), representing a one-dimensional and topological protected bulk mode. The chiral states moving along the boundaries of the sample and the cavity have opposite directions of propagation. This can be understood from the fact that the material is inside the boundary of the sample but outside the boundary of the artificial cavity. In the \textit{End Matter}, we provide more details about this point and also show that inverting the sign of $t_2$ the direction of propagation of both localized modes flips.

\begin{figure}[ht!]
    \centering
    \includegraphics[width=0.9\linewidth]{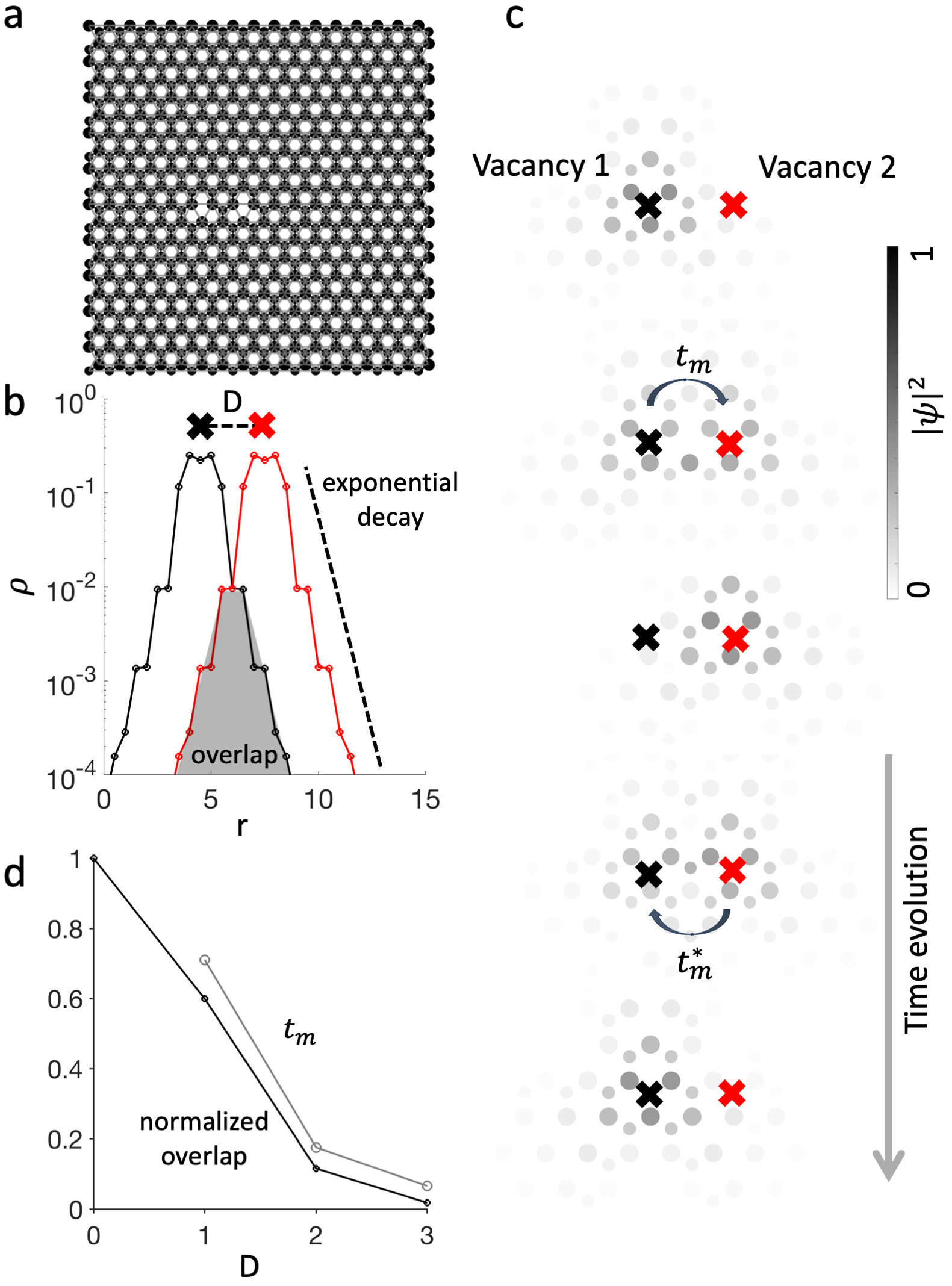}
    \caption{\textbf{(a)} 2D honeycomb lattice with two vacancies placed next to each other at a distance $D$. \textbf{(b)} Angular averaged electron density distribution \(\rho\) nearby the two vacancies (indicated with black crosses). The dashed line shows the exponential decay away from the vacancy cores. \(D\) is the inter-vacancy distance and the shaded region indicates the radial projection of the overlap of the electronic densities. The black curve corresponds to the configuration with only the left vacancy, while the red curve to that with only the right one. \textbf{(c)} Oscillatory dynamics of the electronic density between the two vacancy sites induced by effective hopping ($t_m$ and $t^*_m$). Time evolves from top to bottom; the background color represents the local value of $|\psi|^2$. \textbf{(d)} Overlap of the electronic density (shaded area in panel (c)) and effective hopping parameter \(t_m\) as a function of the distance between the two vacancies ($D$). In the computation, $t_1 = 0.561$ and $t_2 = 0.110$ are used.}
    \label{fig2}
\end{figure}

\color{blue}\textit{Vacancy-induced localized bulk modes} \color{black} -- Having demonstrated the emergence of bulk bound states localized on the lattice imperfections, we are now interested in considering more microscopic structural defects, \textit{e.g.}, vacancies. Vacancies can be readily constructed by removing a single atom in the honeycomb lattice (see Fig. \ref{fig2}(a) for an example involving two nearby vacancies).

In presence of vacancies, the electronic wave-function localizes around the vacancy core. This is demonstrated in Fig.~\ref{fig2}(b) by considering the angular average of the electron wave-function at fixed radial distance, $\rho(r)$. There, we present the distribution of $\rho(r)$ for two different configurations characterized by a single vacancy. Black and red colors correspond respectively to the left and right vacancy in Fig. \ref{fig2}(a). It is shown that $\rho$ displays a maximum at the location of the vacancy and a rapid exponential decay away from it, signaling localization. 

Because of the zero-dimensional nature of the structural defects considered, there is no dynamics associated to the bulk state localized on top of an isolated vacancy. Nevertheless, the situation changes when two or more vacancies are placed next to each other at a distance such that the corresponding electronic wave-functions have a finite overlap, as shown in the case of two vacancies in Fig. \ref{fig2}(c). In this scenario, the overlap of the electronic density (see gray area in Fig.~\ref{fig2}(b)) induces effective interactions between the states localized on the two vacancies, allowing for non-trivial dynamics in the form of hopping between the two vacancy sites. This picture is confirmed in panel (c) of Fig. \ref{fig2} where the electronic density bounces back and forward between the two vacancy sites, inducing an emergence oscillatory behavior. 

We notice that this motion can be understood as the projection of a chiral mode on a finite one-dimensional strip, \textit{akin} to the dynamics of the more standard chiral edge modes on the boundary of the samples. Moreover, we emphasize that this emergence bulk mode is topologically protected meaning that it does not decay in time nor spread in space. On the other hand, its dynamics remain localized along the path designed by the structural defects without any apparent damping. As demonstrated in the attached video animations, this is not the case for a material with no topology, as in the case of $t_2=0$ where the same bulk excitation quickly spreads away from the vacancies and decays in time.


\color{blue}\textit{Effective dynamics} \color{black} -- In order to provide an effective description of the dynamics of the topological states shown in Fig. \ref{fig2}(c), we resort to a simple tight binding model. For simplicity, and with a slight abuse of terminology, we will refer to these topologically localized states as `\textit{vacanons}'. In particular, we propose the following effective Hamiltonian:
\begin{equation}
    \mathcal{H} = \sum_{i,j} t_{m,ij} a_i^\dagger a_j,\label{model}
\end{equation}
where $a_{i}^{\dagger} (a_{i})$ is the creation(annihilation) operator of a vacanon acting on the $i$th vacancy in real space. Moreover, $t_{m,ij}$ is the effective hopping parameter between the $i$th and $j$th vacancies.  

Intuitively, we expect these hopping dynamics to be driven by the quantum tunneling between the vacancy sites; hence, the corresponding effective hopping parameter should appear proportional to the overlap of the electronic densities, namely
\begin{equation}
    t_{m,ij} \propto \int d \boldsymbol{r} \,|\psi(\boldsymbol{r})|^2_{v_1} |\psi(\boldsymbol{r})|^2_{v_2},
\end{equation}
where the label $v_{1,2}$ indicates the wave-function for the configurations with vacancy one and vacancy two. The radial projection of this overlap is shown, in the case of two vacancies (panel (a) in Fig.\ref{fig2}), as a gray region in Fig. \ref{fig2}(b). To confirm the validity of this hypothesis, we compare the overlap of two nearby vacanons and their effective hooping parameter obtained through fitting one eigenenergy appearing within the topological energy gap with Eq.\eqref{model}. The results are shown in Fig.\ref{fig2} (d). It can be seen that the both quantities decreases by increasing the inter-vacancy distance, showing a similar trend and verifying their proportionality.

At this point, the effective hopping parameter is the only parameter necessary to describe the dynamics of the vacanons. More precisely, \(t_m\) is determined by two factors: (i) the inter-vacancy distance as mentioned above and presented in Fig. \ref{fig2}(d), and (ii) the intrinsic properties of material encoded in the microscopic Hamiltonian, Eq. \eqref{simulation}, and the underlying lattice structure. In particular, as shown in \textit{End Matter}, $t_m$ grows monotonically by increasing the electronic imaginary hopping parameter \(t_2\) which determines the topological features.

\begin{figure}[ht!]
    \centering
    \includegraphics[width=\linewidth]{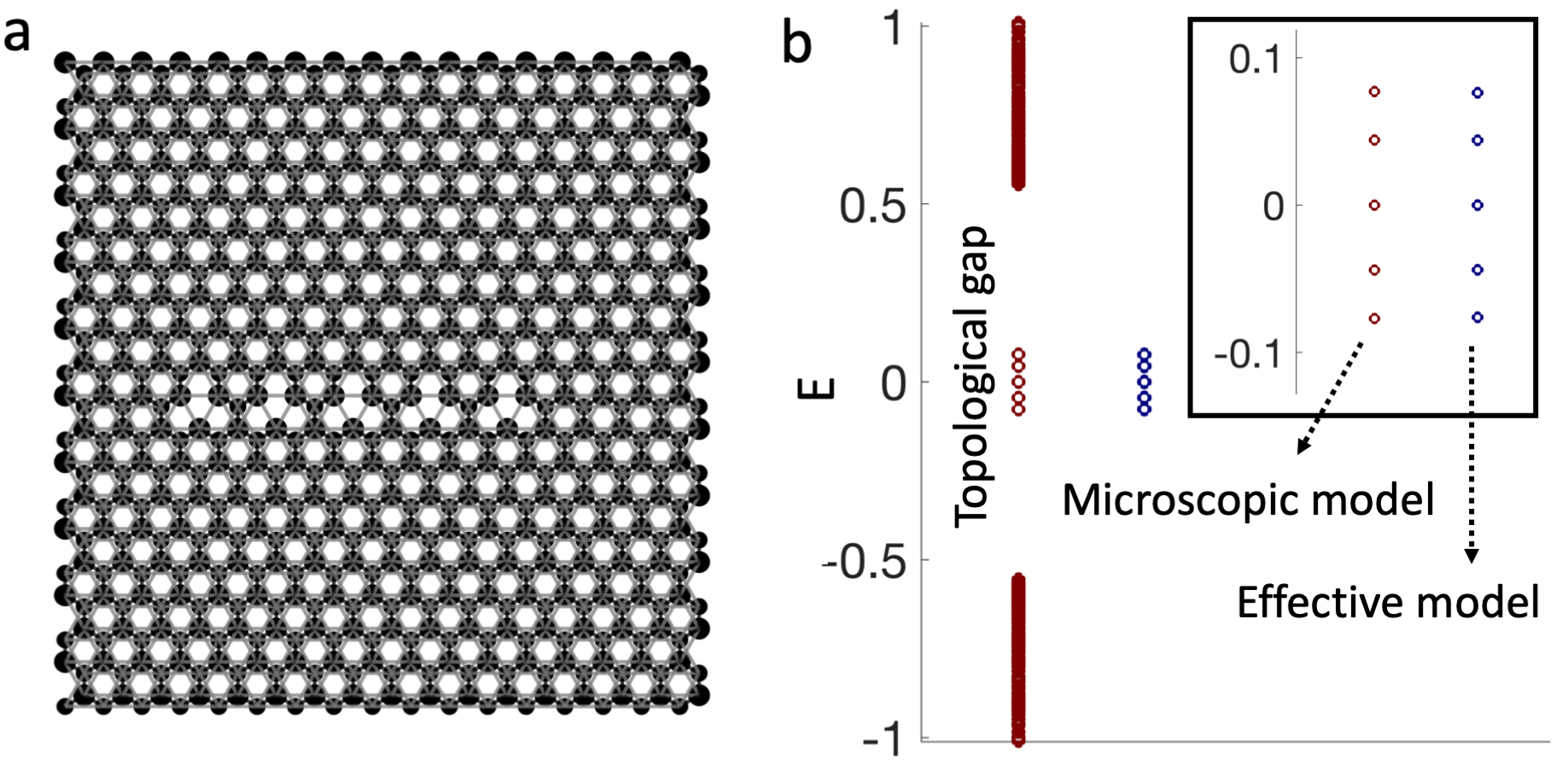}
    \caption{\textbf{(a)} A honeycomb lattice structure with five vacancies arranged along a horizontal strip. \textbf{(b)} Spectrum obtained from the original microscopic Hamiltonian, Eq.\eqref{simulation}, (red symbols) and from the theoretical effective model, Eq.\eqref{model}, (blue symbols). The inset in panel b shows the enlarged view of the states within the topological energy gap induced by the presence of the vacancies.}
    \label{fig3}
\end{figure}

It is important to notice that the emergence of vacancy-induced localized states strongly affects the spectrum of the system with the appearance of electronic states withing the topological gap. We demonstrate this by considering a configuration with five vacancies arranged in a horizontal chain, as presented in Fig. \ref{fig3}(a). We diagonalize the Haldane Hamiltonian, Eq. \eqref{simulation}, in that configuration and show the corresponding electronic spectrum with red symbols in panel (b) of the same figure. Five non-degenerate states clearly emerge within the topological gap. Their non-degenerate nature is due to the interactions between each other induced by the hopping of vacanons.

To establish the validity of our effective model, we also present an estimate for the energy of these states directly using Eq. \eqref{model}, which in this case simplifies to $ \mathcal{H} = \sum_{i=1}^4 t_{m} (a_{i+1}^\dagger a_i + a_{i}^\dagger a_{i+1})$, with $i$ the label of each of the five vacancies. First, we empirically derive the value of the effective hopping parameter $t_m$ by fitting the energy of one of the states with the above effective Hamiltonian. After doing so, no free parameters appear anymore in the effective model and we can predict the rest of the spectrum from it. The results of this procedure are shown with blue symbols in panel (b) of Fig. \ref{fig3}. It is evident that this simple effective model provides a very good estimate for the energy of the vacancy-induced localized states that emerge within the topological gap. In the \textit{End Matter}, we provide another successful test by increasing the distance between the vacancies. From there, we can also notice that by increasing the distance, the effective hopping parameter decreases and the energies of the localized states become closer to each other. This is reasonable since the degeneracy is lifted by the hopping, and when the latter vanishes we do expect all the localized states to become degenerate.

\begin{figure}[ht!]
    \centering
    \includegraphics[width=\linewidth]{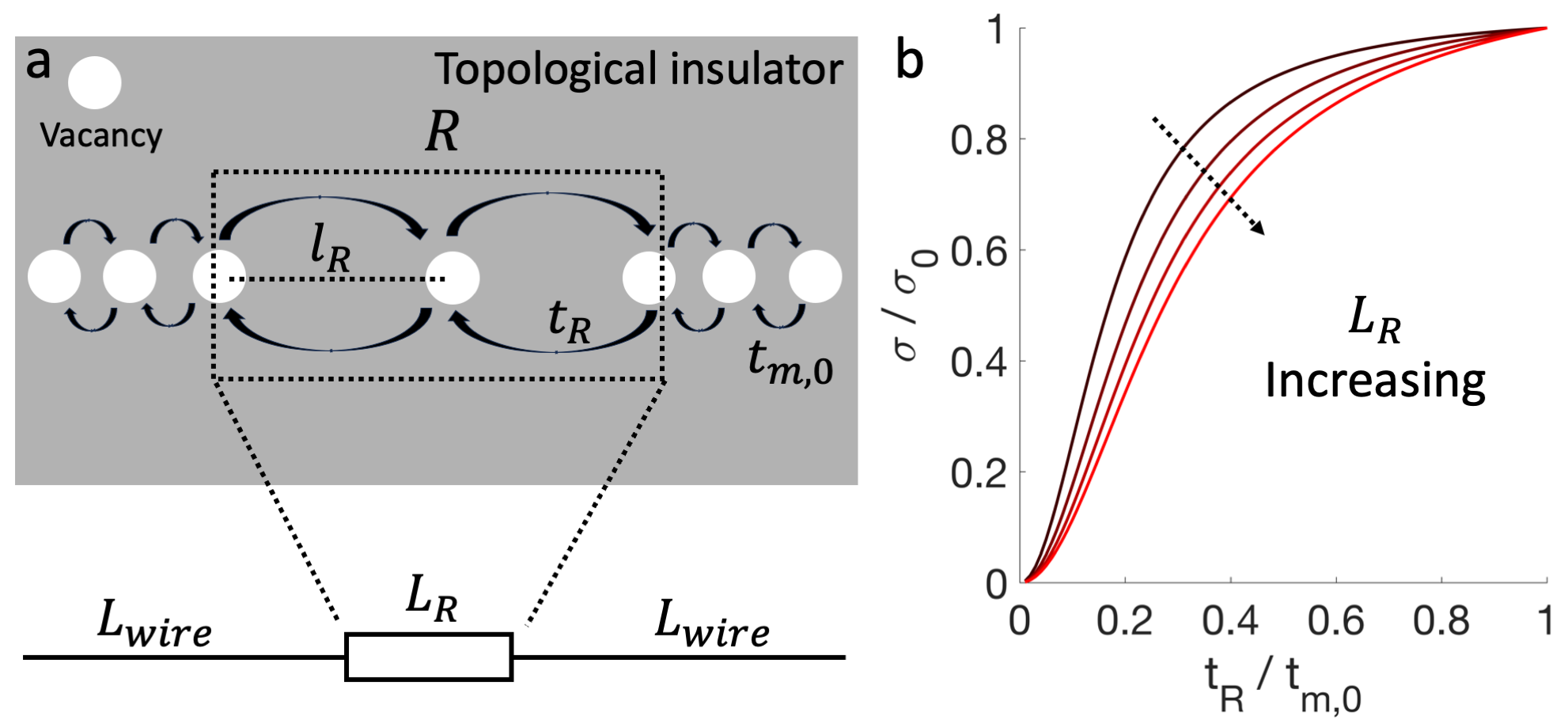}
    \caption{\textbf{(a)} A cartoon of the design of an atomic-scale resistive circuit achieved by controlling the inter-vacancy distance in a TI. \textbf{(b)} Overall conductivity \(\sigma\) of the resistive circuit with resistor's effective hopping parameter \(t_R\) and for different lengths of the resistor. From black to red curves, the length of the resistor is \(L_{R}=\) 1, 3, 5, 7 resistor's inter-vacancy distance \(l_R\). \(\sigma_0\) is the conductivity of the circuit without the resistor. In the computation, the length of the wire is fixed to \(L_{wire} = 20\) wire's inter-vacancy distance, the Fermi energy \(\epsilon_F = 0\) and the inelastic scattering parameter \(\eta = t_{m,0}\).}
    \label{fig4}
\end{figure}

\color{blue}\textit{Atomic-scale resistive circuit} \color{black} -- Based on the vacanon dynamics described by the effective tight binding model in Eq.\eqref{model}, we propose the design of an atomic-scale resistive circuit. The electric wire can be built from a chain of vacancies with small inter-vacancy distance, while a resistor from a subset of vacancies with larger inter-vacancy distance. This idea is presented in Fig.\ref{fig4} (a) and it is based on the observation that a larger inter-vacancy distance implies a lower hopping parameter and, hence, inhibits the mobility of the vacanon states. Here, we define the length of the wire and the length of the resistor as $L_{\text{wire}}$ and $L_R$.

Since these states hopping from one vacancy site to another are microscopically composed of electrons, they carry a finite electric charge $e_v$. Using the effective Hamiltonian in Eq.\eqref{model}, the conductivity of a wire of vacancies can be obtained using Kubo formula, \cite{PhysRevLett.129.196601,PhysRevLett.115.106601}
\begin{equation}
    \sigma_{\text{wire}} =  \dfrac{2 \hbar e_v^2}{\pi L_{\text{wire}}}  \mathrm{Tr}\left[ \mathrm{Im}\mathcal{G}(\epsilon_F + i \eta) v \mathrm{Im}\mathcal{G}(\epsilon_F - i \eta) v \right].
\end{equation}
Here $L_{\text{wire}}$ is length of wire, $\epsilon_F$ is Fermi energy, $\eta$ is the inelastic scattering parameter and $v = -(i/\hbar) [X,\mathcal{H}]$ is the velocity operator along the wire with $X$ the position operator of the vacancies. The propagator $\mathcal{G}(\epsilon_F + i \eta)$ is given by solving $\left[(\epsilon_F + i \eta) I - \mathcal{H}\right]\mathcal{G} = I$ with $I$ the identity matrix. Similarly the conductivity of the resistor \(\sigma_{R}\) can be obtained from the same effective Hamiltonian but with smaller effective hopping parameter \(t_R < t_m\), that is achieved controlling the inter-vacancy distance.

The fundamental element of a resistive circuit is a wire-resistor-wire chain (see Fig.\ref{fig4} (a)). This can be then built by arranging a set of vacancies along a horizontal strip and by controlling their distances, namely making the latter larger in an intermediate interval of length $L_R$. The overall conductivity \(\sigma\) of such a resistive circuit can be obtained according to the sum rule of resistances in series,
\begin{equation}
    \sigma = \left( \dfrac{2 \sigma_{\text{wire}}^{-1}L_{\text{wire}} + \sigma_{R}^{-1}L_{R}}{ 2 L_{\text{wire}} + L_{R}}\right)^{-1}.
\end{equation}

Fig.\ref{fig4} (b) show the predicted overall conductivity \(\sigma\) in function of the ratio between the resistor hopping parameter \(t_R\) and the wire hopping parameter $t_{m,0}$ for different resistor size $L_R$. The conductivity is normalized by the wire conductivity $\sigma_0$. When the resistor's effective hopping parameter \(t_R\) is zero, the circuit is open and the overall conductivity vanishes. On the other hand, when the effective hopping parameter is uniform along the circuit \(t_R = t_{m,0}\), there is no resistor anymore and the overall conductivity coincides with that of the wire $\sigma_0$. In addition to that, the length of the resistor $L_R$ has also an effect on the overall conductivity as shown in Fig.\ref{fig4} (b) from black to red curves. As expected from basic arguments, a longer resistor implies a smaller overall conductivity.

\color{blue}\textit{Conclusions} \color{black} -- Recent advances in controlling defects in two-dimensional materials at the atomic scale have opened new avenues for the development of atomic-scale electronic devices \cite{PhysRevLett.107.116803,Vicarelli2015,Robertson2012}. Due to their bounded and localized nature around defects, topological states emerge as promising candidates for designing such devices.

In this work, we investigate the dynamics of edge states induced by vacancies, which we refer to as `\textit{vacanons},' by simulating the electronic properties of a topological insulator within the Haldane model framework. We find that pairs of vacancies in a topological insulator can interact through the hopping of vacanons between them. The hopping amplitude is shown to be directly correlated with the overlap of the electronic densities localized around the vacancies. To quantitatively describe vacanon dynamics, we develop an effective tight-binding Hamiltonian, where the hopping parameters depend on both the inter-vacancy distance and the intrinsic material properties. Finally, we propose that vacanons offer a novel platform for designing atomic-scale resistive circuits, as their dynamics can be fully determined from the configuration and the effective Hamiltonian.

Our results demonstrate that the dynamics of vacancy-induced excitations in topological insulators can be effectively controlled using currently available technologies, highlighting their promising potential for applications in atomic-scale electronic devices. Based on the impressive advancements in defect engineering, such as vacancy control, in 2D materials
\cite{Mao2016,doi:10.34133/2019/4641739}, we anticipate that the future realization of our designed resistive circuits might be already within reach.

\color{blue}{\it Acknowledgments} \color{black} -- The authors acknowledge the support of the Shanghai Municipal Science and Technology Major Project (Grant No. 2019SHZDZX01). MB acknowledges the support of the sponsorship from the Yangyang Development Fund.

\clearpage
\newpage
\section*{End Matter}\label{end}
\begin{figure}[ht!]
    \centering
    \includegraphics[width=0.9\linewidth]{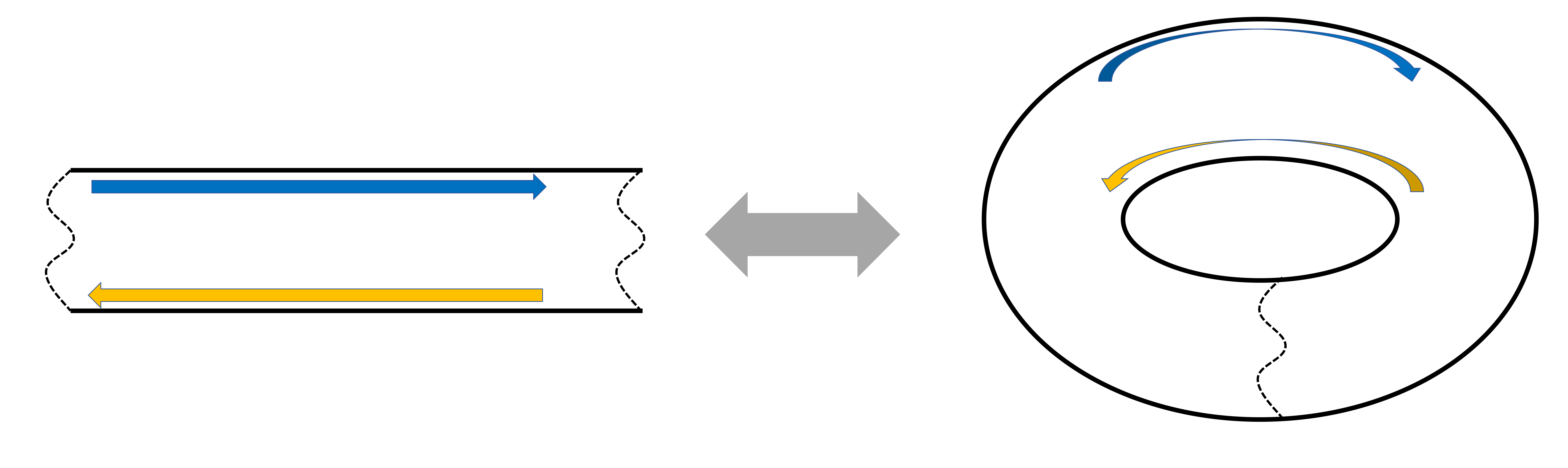}
    \caption{An illustration of chiral states on a strip and on a donut. A strip of the material can be deformed and connected continuously to a donut, and a donut can be torn and deformed into a strip back. The arrows indicate the direction of the chiral states when $C=-1$.}
    \label{fig5}
\end{figure}

\begin{figure}[ht!]
    \centering
    \includegraphics[width=\linewidth]{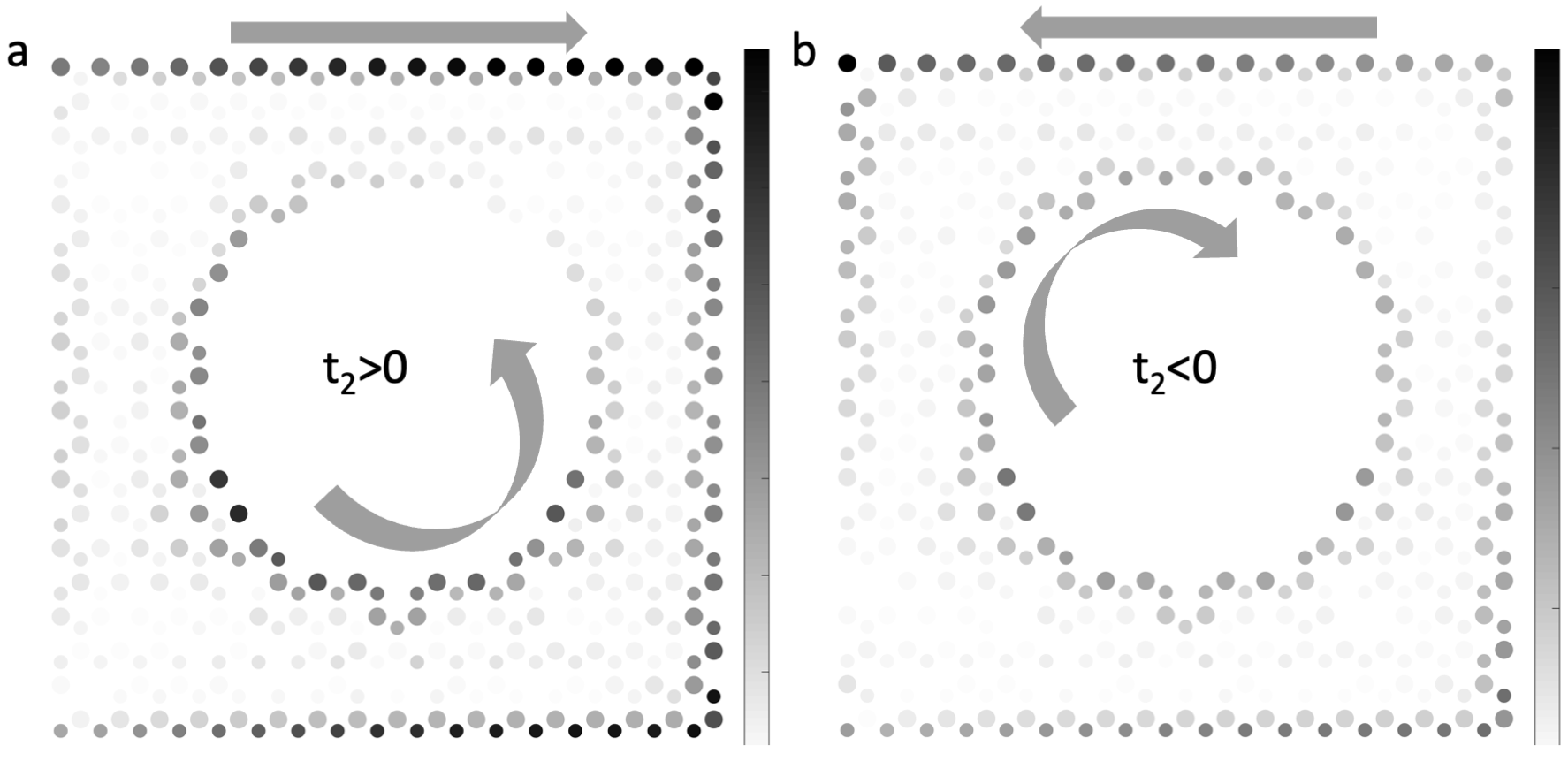}
    \caption{The chiral propagation of the topological edge modes for opposite signs of the imaginary hopping \(t_2\) in the Haldane model, Eq. \eqref{simulation}. As the gray arrows indicate, the chirality of the modes is evidently opposite.}
    \label{fig6}
\end{figure}

{\color{blue} \it Appendix A: Chirality of topological modes -} 
In order to understand the chirality of the topological states moving along the boundary and the cavity (see Fig. \ref{fig1}(d)) in more detail, we provide a cartoon of the material both with and without the cavity, as shown in Fig.~\ref{fig5}. The state along the lower edge of the strip (yellow) corresponds to the chiral state along the hole of the donut, while the state along the upper edge (blue) corresponds to the chiral state along the boundary of the donut. It is clear that both states are determined by the Chern number. The donut can be continuously deformed into the shape as shown in the panels(c)(d) of Fig.\ref{fig1} without changing the chirality of states since the latter is topologically protected. 
When the Chern number changes sign, all topological localized states reverse their chirality. This can be directly controlled in the Haldane mode, Eq. \eqref{simulation}, by changing the sign of imaginary hopping \(t_2\), as demonstrated in Fig. \ref{fig6}.

\begin{figure}[ht!]
    \centering
    \includegraphics[width=\linewidth]{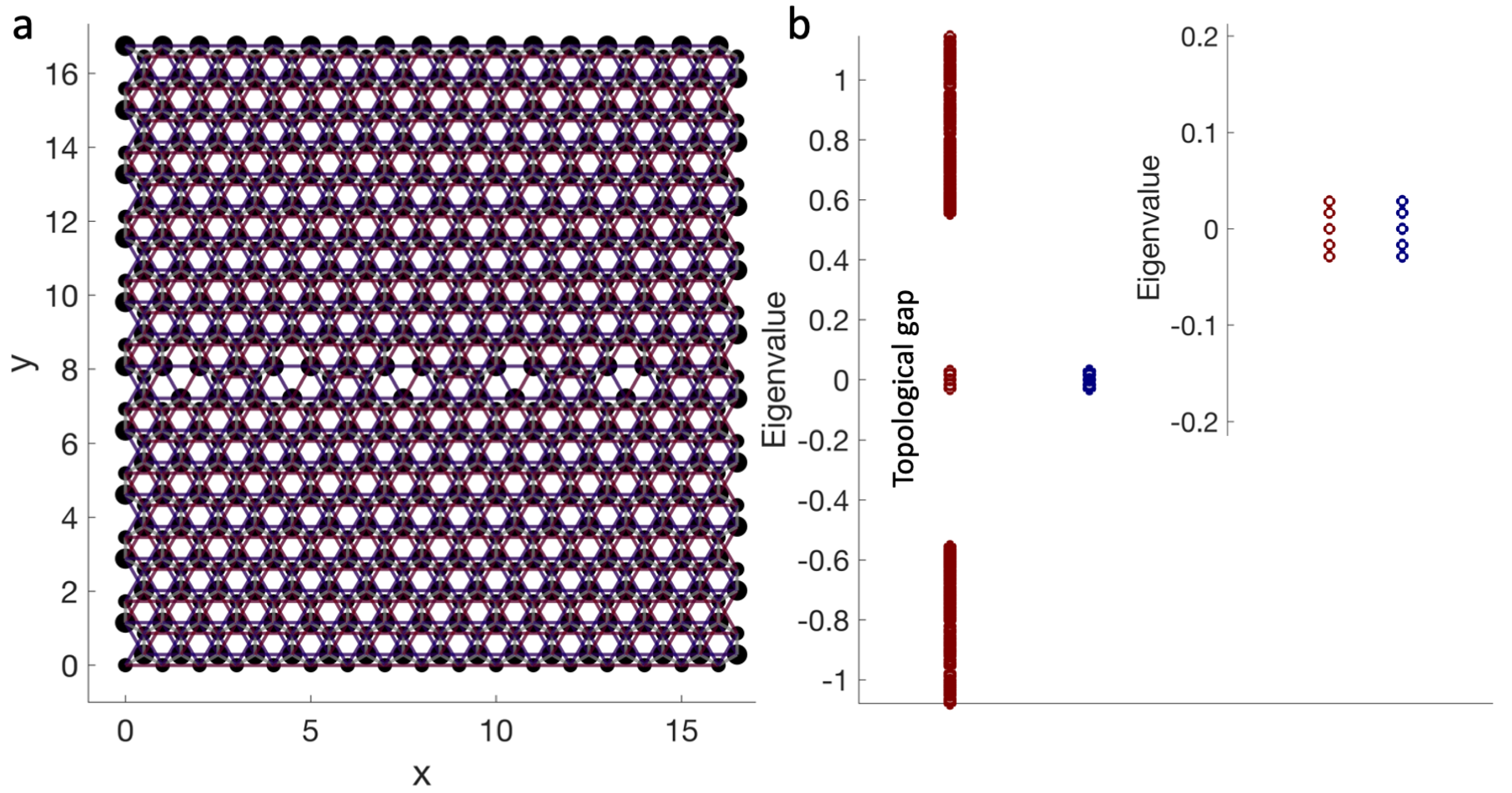}
    \caption{\textbf{(a)} Honeycomb lattice with an array of five vacancies. \textbf{(b)} Electronic spectrum obtained from the microscopic Haldane model Eq.\eqref{simulation} (red) and the effective model Eq.\eqref{model} (blue).}
    \label{fig7}
\end{figure}

{\it \color{blue} Appendix B: Vacancies with larger distance -} In this Appendix, we provide further analysis to confirm the validity of our effective model, Eq. \eqref{model}, by considering an array of vacancies with larger inter-vacancy distance with respect to the case presented in the main text. The structure of this sample is shown in Fig. \ref{fig7}(a), while the spectrum computed both with the original microscopic Haldane model and the effective vacanon Hamiltonian is presented in panel (b) of the same Figure. The agreement is excellent, confirming the validity of our theoretical model.

\begin{figure}[ht!]
    \centering
    \includegraphics[width=0.7\linewidth]{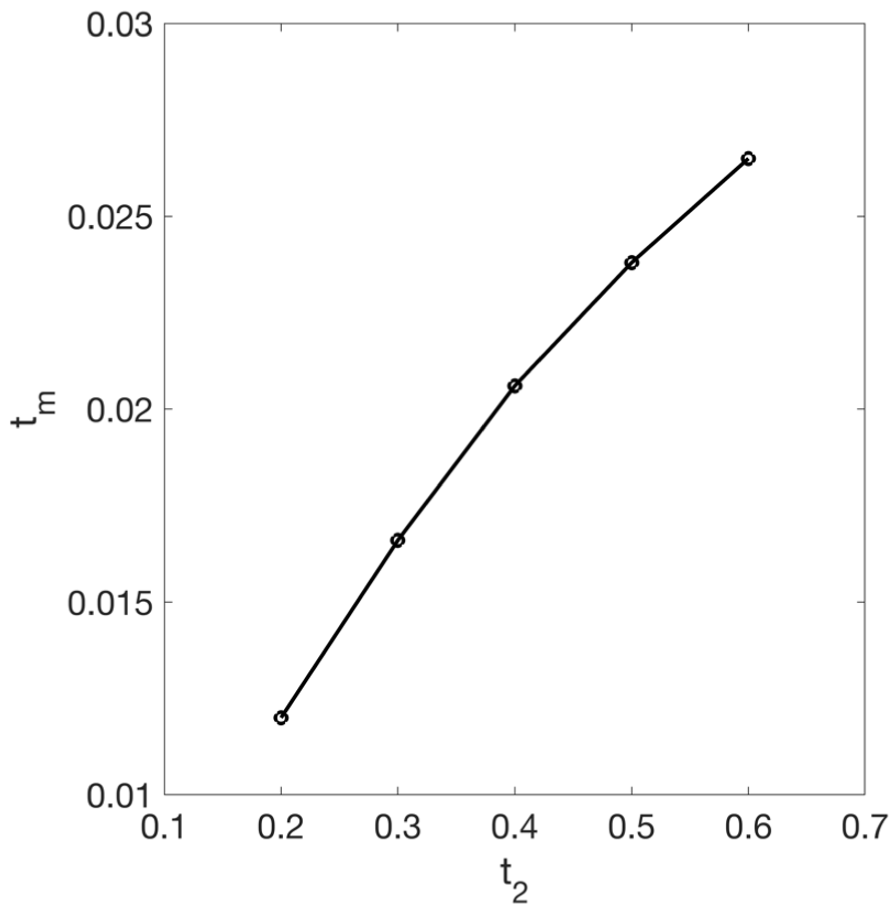}
    \caption{Effective vacanon hopping parameter $t_m$ as a function of the NNN hopping in the Haldane microscopic model $t_2$.}
    \label{fig8}
\end{figure}

{\it \color{blue} Appendix C: Extended analysis of the effective hopping parameter -}
In Fig. \ref{fig8}, we present the dependence of the effective hopping parameter \(t_m\) on the  imaginary hopping parameter \(t_2\) in the Haldane Hamiltonian, Eq. \eqref{simulation}. The vacanon hopping increases with $t_2$.

\begin{thebibliography}{43}%
\makeatletter
\providecommand \@ifxundefined [1]{%
 \@ifx{#1\undefined}
}%
\providecommand \@ifnum [1]{%
 \ifnum #1\expandafter \@firstoftwo
 \else \expandafter \@secondoftwo
 \fi
}%
\providecommand \@ifx [1]{%
 \ifx #1\expandafter \@firstoftwo
 \else \expandafter \@secondoftwo
 \fi
}%
\providecommand \natexlab [1]{#1}%
\providecommand \enquote  [1]{``#1''}%
\providecommand \bibnamefont  [1]{#1}%
\providecommand \bibfnamefont [1]{#1}%
\providecommand \citenamefont [1]{#1}%
\providecommand \href@noop [0]{\@secondoftwo}%
\providecommand \href [0]{\begingroup \@sanitize@url \@href}%
\providecommand \@href[1]{\@@startlink{#1}\@@href}%
\providecommand \@@href[1]{\endgroup#1\@@endlink}%
\providecommand \@sanitize@url [0]{\catcode `\\12\catcode `\$12\catcode `\&12\catcode `\#12\catcode `\^12\catcode `\_12\catcode `\%12\relax}%
\providecommand \@@startlink[1]{}%
\providecommand \@@endlink[0]{}%
\providecommand \url  [0]{\begingroup\@sanitize@url \@url }%
\providecommand \@url [1]{\endgroup\@href {#1}{\urlprefix }}%
\providecommand \urlprefix  [0]{URL }%
\providecommand \Eprint [0]{\href }%
\providecommand \doibase [0]{http://dx.doi.org/}%
\providecommand \selectlanguage [0]{\@gobble}%
\providecommand \bibinfo  [0]{\@secondoftwo}%
\providecommand \bibfield  [0]{\@secondoftwo}%
\providecommand \translation [1]{[#1]}%
\providecommand \BibitemOpen [0]{}%
\providecommand \bibitemStop [0]{}%
\providecommand \bibitemNoStop [0]{.\EOS\space}%
\providecommand \EOS [0]{\spacefactor3000\relax}%
\providecommand \BibitemShut  [1]{\csname bibitem#1\endcsname}%
\let\auto@bib@innerbib\@empty
\bibitem [{\citenamefont {Hasan}\ and\ \citenamefont {Kane}(2010)}]{RevModPhys.82.3045}%
  \BibitemOpen
  \bibfield  {author} {\bibinfo {author} {\bibfnamefont {M.~Z.}\ \bibnamefont {Hasan}}\ and\ \bibinfo {author} {\bibfnamefont {C.~L.}\ \bibnamefont {Kane}},\ }\href {\doibase 10.1103/RevModPhys.82.3045} {\bibfield  {journal} {\bibinfo  {journal} {Rev. Mod. Phys.}\ }\textbf {\bibinfo {volume} {82}},\ \bibinfo {pages} {3045} (\bibinfo {year} {2010})}\BibitemShut {NoStop}%
\bibitem [{\citenamefont {Rachel}(2018)}]{Rachel_2018}%
  \BibitemOpen
  \bibfield  {author} {\bibinfo {author} {\bibfnamefont {S.}~\bibnamefont {Rachel}},\ }\href {\doibase 10.1088/1361-6633/aad6a6} {\bibfield  {journal} {\bibinfo  {journal} {Reports on Progress in Physics}\ }\textbf {\bibinfo {volume} {81}},\ \bibinfo {pages} {116501} (\bibinfo {year} {2018})}\BibitemShut {NoStop}%
\bibitem [{\citenamefont {Hasan}\ and\ \citenamefont {Moore}(2011)}]{annurev:/content/journals/10.1146/annurev-conmatphys-062910-140432}%
  \BibitemOpen
  \bibfield  {author} {\bibinfo {author} {\bibfnamefont {M.~Z.}\ \bibnamefont {Hasan}}\ and\ \bibinfo {author} {\bibfnamefont {J.~E.}\ \bibnamefont {Moore}},\ }\href {\doibase https://doi.org/10.1146/annurev-conmatphys-062910-140432} {\bibfield  {journal} {\bibinfo  {journal} {Annual Review of Condensed Matter Physics}\ }\textbf {\bibinfo {volume} {2}},\ \bibinfo {pages} {55} (\bibinfo {year} {2011})}\BibitemShut {NoStop}%
\bibitem [{\citenamefont {Moore}(2010)}]{Moore2010}%
  \BibitemOpen
  \bibfield  {author} {\bibinfo {author} {\bibfnamefont {J.~E.}\ \bibnamefont {Moore}},\ }\href {\doibase 10.1038/nature08916} {\bibfield  {journal} {\bibinfo  {journal} {Nature}\ }\textbf {\bibinfo {volume} {464}},\ \bibinfo {pages} {194} (\bibinfo {year} {2010})}\BibitemShut {NoStop}%
\bibitem [{\citenamefont {Kane}\ and\ \citenamefont {Mele}(2005{\natexlab{a}})}]{PhysRevLett.95.146802}%
  \BibitemOpen
  \bibfield  {author} {\bibinfo {author} {\bibfnamefont {C.~L.}\ \bibnamefont {Kane}}\ and\ \bibinfo {author} {\bibfnamefont {E.~J.}\ \bibnamefont {Mele}},\ }\href {\doibase 10.1103/PhysRevLett.95.146802} {\bibfield  {journal} {\bibinfo  {journal} {Phys. Rev. Lett.}\ }\textbf {\bibinfo {volume} {95}},\ \bibinfo {pages} {146802} (\bibinfo {year} {2005}{\natexlab{a}})}\BibitemShut {NoStop}%
\bibitem [{\citenamefont {Kane}\ and\ \citenamefont {Mele}(2005{\natexlab{b}})}]{PhysRevLett.95.226801}%
  \BibitemOpen
  \bibfield  {author} {\bibinfo {author} {\bibfnamefont {C.~L.}\ \bibnamefont {Kane}}\ and\ \bibinfo {author} {\bibfnamefont {E.~J.}\ \bibnamefont {Mele}},\ }\href {\doibase 10.1103/PhysRevLett.95.226801} {\bibfield  {journal} {\bibinfo  {journal} {Phys. Rev. Lett.}\ }\textbf {\bibinfo {volume} {95}},\ \bibinfo {pages} {226801} (\bibinfo {year} {2005}{\natexlab{b}})}\BibitemShut {NoStop}%
\bibitem [{\citenamefont {Bernevig}\ and\ \citenamefont {Zhang}(2006)}]{PhysRevLett.96.106802}%
  \BibitemOpen
  \bibfield  {author} {\bibinfo {author} {\bibfnamefont {B.~A.}\ \bibnamefont {Bernevig}}\ and\ \bibinfo {author} {\bibfnamefont {S.-C.}\ \bibnamefont {Zhang}},\ }\href {\doibase 10.1103/PhysRevLett.96.106802} {\bibfield  {journal} {\bibinfo  {journal} {Phys. Rev. Lett.}\ }\textbf {\bibinfo {volume} {96}},\ \bibinfo {pages} {106802} (\bibinfo {year} {2006})}\BibitemShut {NoStop}%
\bibitem [{\citenamefont {Haldane}(1988)}]{PhysRevLett.61.2015}%
  \BibitemOpen
  \bibfield  {author} {\bibinfo {author} {\bibfnamefont {F.~D.~M.}\ \bibnamefont {Haldane}},\ }\href {\doibase 10.1103/PhysRevLett.61.2015} {\bibfield  {journal} {\bibinfo  {journal} {Phys. Rev. Lett.}\ }\textbf {\bibinfo {volume} {61}},\ \bibinfo {pages} {2015} (\bibinfo {year} {1988})}\BibitemShut {NoStop}%
\bibitem [{\citenamefont {Xu}\ \emph {et~al.}(2014)\citenamefont {Xu}, \citenamefont {Miotkowski}, \citenamefont {Liu}, \citenamefont {Tian}, \citenamefont {Nam}, \citenamefont {Alidoust}, \citenamefont {Hu}, \citenamefont {Shih}, \citenamefont {Hasan},\ and\ \citenamefont {Chen}}]{Xu2014}%
  \BibitemOpen
  \bibfield  {author} {\bibinfo {author} {\bibfnamefont {Y.}~\bibnamefont {Xu}}, \bibinfo {author} {\bibfnamefont {I.}~\bibnamefont {Miotkowski}}, \bibinfo {author} {\bibfnamefont {C.}~\bibnamefont {Liu}}, \bibinfo {author} {\bibfnamefont {J.}~\bibnamefont {Tian}}, \bibinfo {author} {\bibfnamefont {H.}~\bibnamefont {Nam}}, \bibinfo {author} {\bibfnamefont {N.}~\bibnamefont {Alidoust}}, \bibinfo {author} {\bibfnamefont {J.}~\bibnamefont {Hu}}, \bibinfo {author} {\bibfnamefont {C.-K.}\ \bibnamefont {Shih}}, \bibinfo {author} {\bibfnamefont {M.~Z.}\ \bibnamefont {Hasan}}, \ and\ \bibinfo {author} {\bibfnamefont {Y.~P.}\ \bibnamefont {Chen}},\ }\href {\doibase 10.1038/nphys3140} {\bibfield  {journal} {\bibinfo  {journal} {Nature Physics}\ }\textbf {\bibinfo {volume} {10}},\ \bibinfo {pages} {956} (\bibinfo {year} {2014})}\BibitemShut {NoStop}%
\bibitem [{\citenamefont {Breunig}\ and\ \citenamefont {Ando}(2022)}]{Breunig2022}%
  \BibitemOpen
  \bibfield  {author} {\bibinfo {author} {\bibfnamefont {O.}~\bibnamefont {Breunig}}\ and\ \bibinfo {author} {\bibfnamefont {Y.}~\bibnamefont {Ando}},\ }\href {\doibase 10.1038/s42254-021-00402-6} {\bibfield  {journal} {\bibinfo  {journal} {Nature Reviews Physics}\ }\textbf {\bibinfo {volume} {4}},\ \bibinfo {pages} {184} (\bibinfo {year} {2022})}\BibitemShut {NoStop}%
\bibitem [{\citenamefont {Xiong}\ \emph {et~al.}(2021)\citenamefont {Xiong}, \citenamefont {Zhong}, \citenamefont {Wang},\ and\ \citenamefont {Li}}]{Xiong2021-oz}%
  \BibitemOpen
  \bibfield  {author} {\bibinfo {author} {\bibfnamefont {Z.}~\bibnamefont {Xiong}}, \bibinfo {author} {\bibfnamefont {L.}~\bibnamefont {Zhong}}, \bibinfo {author} {\bibfnamefont {H.}~\bibnamefont {Wang}}, \ and\ \bibinfo {author} {\bibfnamefont {X.}~\bibnamefont {Li}},\ }\href@noop {} {\bibfield  {journal} {\bibinfo  {journal} {Materials (Basel)}\ }\textbf {\bibinfo {volume} {14}} (\bibinfo {year} {2021})}\BibitemShut {NoStop}%
\bibitem [{\citenamefont {Science}(1998)}]{doi:10.1126/science.281.5379.939}%
  \BibitemOpen
  \bibfield  {author} {\bibinfo {author} {\bibnamefont {Science}},\ }\href {\doibase 10.1126/science.281.5379.939} {\bibfield  {journal} {\bibinfo  {journal} {Science}\ }\textbf {\bibinfo {volume} {281}},\ \bibinfo {pages} {939} (\bibinfo {year} {1998})}\BibitemShut {NoStop}%
\bibitem [{\citenamefont {Zhang}\ \emph {et~al.}(2023)\citenamefont {Zhang}, \citenamefont {Kang},\ and\ \citenamefont {Wei}}]{Zhang2023}%
  \BibitemOpen
  \bibfield  {author} {\bibinfo {author} {\bibfnamefont {X.}~\bibnamefont {Zhang}}, \bibinfo {author} {\bibfnamefont {J.}~\bibnamefont {Kang}}, \ and\ \bibinfo {author} {\bibfnamefont {S.-H.}\ \bibnamefont {Wei}},\ }\href {\doibase 10.1038/s43588-023-00403-8} {\bibfield  {journal} {\bibinfo  {journal} {Nature Computational Science}\ }\textbf {\bibinfo {volume} {3}},\ \bibinfo {pages} {210} (\bibinfo {year} {2023})}\BibitemShut {NoStop}%
\bibitem [{\citenamefont {Tuller}\ and\ \citenamefont {Bishop}(2010)}]{10.1246/cl.2010.1226}%
  \BibitemOpen
  \bibfield  {author} {\bibinfo {author} {\bibfnamefont {H.~L.}\ \bibnamefont {Tuller}}\ and\ \bibinfo {author} {\bibfnamefont {S.~R.}\ \bibnamefont {Bishop}},\ }\href {\doibase 10.1246/cl.2010.1226} {\bibfield  {journal} {\bibinfo  {journal} {Chemistry Letters}\ }\textbf {\bibinfo {volume} {39}},\ \bibinfo {pages} {1226} (\bibinfo {year} {2010})}\BibitemShut {NoStop}%
\bibitem [{\citenamefont {de~Juan}\ \emph {et~al.}(2014)\citenamefont {de~Juan}, \citenamefont {R\"uegg},\ and\ \citenamefont {Lee}}]{PhysRevB.89.161117}%
  \BibitemOpen
  \bibfield  {author} {\bibinfo {author} {\bibfnamefont {F.}~\bibnamefont {de~Juan}}, \bibinfo {author} {\bibfnamefont {A.}~\bibnamefont {R\"uegg}}, \ and\ \bibinfo {author} {\bibfnamefont {D.-H.}\ \bibnamefont {Lee}},\ }\href {\doibase 10.1103/PhysRevB.89.161117} {\bibfield  {journal} {\bibinfo  {journal} {Phys. Rev. B}\ }\textbf {\bibinfo {volume} {89}},\ \bibinfo {pages} {161117} (\bibinfo {year} {2014})}\BibitemShut {NoStop}%
\bibitem [{\citenamefont {Lu}\ \emph {et~al.}(2011)\citenamefont {Lu}, \citenamefont {Shan}, \citenamefont {Lu},\ and\ \citenamefont {Shen}}]{Lu2011}%
  \BibitemOpen
  \bibfield  {author} {\bibinfo {author} {\bibfnamefont {J.}~\bibnamefont {Lu}}, \bibinfo {author} {\bibfnamefont {W.-Y.}\ \bibnamefont {Shan}}, \bibinfo {author} {\bibfnamefont {H.-Z.}\ \bibnamefont {Lu}}, \ and\ \bibinfo {author} {\bibfnamefont {S.-Q.}\ \bibnamefont {Shen}},\ }\href {\doibase 10.1088/1367-2630/13/10/103016} {\bibfield  {journal} {\bibinfo  {journal} {New Journal of Physics}\ }\textbf {\bibinfo {volume} {13}},\ \bibinfo {pages} {103016} (\bibinfo {year} {2011})}\BibitemShut {NoStop}%
\bibitem [{\citenamefont {Kao}\ \emph {et~al.}(2021)\citenamefont {Kao}, \citenamefont {Knolle}, \citenamefont {Hal\'asz}, \citenamefont {Moessner},\ and\ \citenamefont {Perkins}}]{PhysRevX.11.011034}%
  \BibitemOpen
  \bibfield  {author} {\bibinfo {author} {\bibfnamefont {W.-H.}\ \bibnamefont {Kao}}, \bibinfo {author} {\bibfnamefont {J.}~\bibnamefont {Knolle}}, \bibinfo {author} {\bibfnamefont {G.~B.}\ \bibnamefont {Hal\'asz}}, \bibinfo {author} {\bibfnamefont {R.}~\bibnamefont {Moessner}}, \ and\ \bibinfo {author} {\bibfnamefont {N.~B.}\ \bibnamefont {Perkins}},\ }\href {\doibase 10.1103/PhysRevX.11.011034} {\bibfield  {journal} {\bibinfo  {journal} {Phys. Rev. X}\ }\textbf {\bibinfo {volume} {11}},\ \bibinfo {pages} {011034} (\bibinfo {year} {2021})}\BibitemShut {NoStop}%
\bibitem [{\citenamefont {Sablikov}\ and\ \citenamefont {Sukhanov}(2015)}]{PhysRevB.91.075412}%
  \BibitemOpen
  \bibfield  {author} {\bibinfo {author} {\bibfnamefont {V.~A.}\ \bibnamefont {Sablikov}}\ and\ \bibinfo {author} {\bibfnamefont {A.~A.}\ \bibnamefont {Sukhanov}},\ }\href {\doibase 10.1103/PhysRevB.91.075412} {\bibfield  {journal} {\bibinfo  {journal} {Phys. Rev. B}\ }\textbf {\bibinfo {volume} {91}},\ \bibinfo {pages} {075412} (\bibinfo {year} {2015})}\BibitemShut {NoStop}%
\bibitem [{\citenamefont {Shan}\ \emph {et~al.}(2011)\citenamefont {Shan}, \citenamefont {Lu}, \citenamefont {Lu},\ and\ \citenamefont {Shen}}]{PhysRevB.84.035307}%
  \BibitemOpen
  \bibfield  {author} {\bibinfo {author} {\bibfnamefont {W.-Y.}\ \bibnamefont {Shan}}, \bibinfo {author} {\bibfnamefont {J.}~\bibnamefont {Lu}}, \bibinfo {author} {\bibfnamefont {H.-Z.}\ \bibnamefont {Lu}}, \ and\ \bibinfo {author} {\bibfnamefont {S.-Q.}\ \bibnamefont {Shen}},\ }\href {\doibase 10.1103/PhysRevB.84.035307} {\bibfield  {journal} {\bibinfo  {journal} {Phys. Rev. B}\ }\textbf {\bibinfo {volume} {84}},\ \bibinfo {pages} {035307} (\bibinfo {year} {2011})}\BibitemShut {NoStop}%
\bibitem [{\citenamefont {Lin}\ \emph {et~al.}(2023)\citenamefont {Lin}, \citenamefont {Wang}, \citenamefont {Liu}, \citenamefont {Xue}, \citenamefont {Zhang}, \citenamefont {Chong},\ and\ \citenamefont {Jiang}}]{Lin2023}%
  \BibitemOpen
  \bibfield  {author} {\bibinfo {author} {\bibfnamefont {Z.-K.}\ \bibnamefont {Lin}}, \bibinfo {author} {\bibfnamefont {Q.}~\bibnamefont {Wang}}, \bibinfo {author} {\bibfnamefont {Y.}~\bibnamefont {Liu}}, \bibinfo {author} {\bibfnamefont {H.}~\bibnamefont {Xue}}, \bibinfo {author} {\bibfnamefont {B.}~\bibnamefont {Zhang}}, \bibinfo {author} {\bibfnamefont {Y.}~\bibnamefont {Chong}}, \ and\ \bibinfo {author} {\bibfnamefont {J.-H.}\ \bibnamefont {Jiang}},\ }\href {https://doi.org/10.1038/s42254-023-00602-2} {\bibfield  {journal} {\bibinfo  {journal} {Nature Reviews Physics}\ }\textbf {\bibinfo {volume} {5}},\ \bibinfo {pages} {483} (\bibinfo {year} {2023})}\BibitemShut {NoStop}%
\bibitem [{\citenamefont {Sun}\ \emph {et~al.}(2021)\citenamefont {Sun}, \citenamefont {Zhu},\ and\ \citenamefont {Hughes}}]{PhysRevLett.127.066401}%
  \BibitemOpen
  \bibfield  {author} {\bibinfo {author} {\bibfnamefont {X.-Q.}\ \bibnamefont {Sun}}, \bibinfo {author} {\bibfnamefont {P.}~\bibnamefont {Zhu}}, \ and\ \bibinfo {author} {\bibfnamefont {T.~L.}\ \bibnamefont {Hughes}},\ }\href {\doibase 10.1103/PhysRevLett.127.066401} {\bibfield  {journal} {\bibinfo  {journal} {Phys. Rev. Lett.}\ }\textbf {\bibinfo {volume} {127}},\ \bibinfo {pages} {066401} (\bibinfo {year} {2021})}\BibitemShut {NoStop}%
\bibitem [{\citenamefont {R\"uegg}\ \emph {et~al.}(2013)\citenamefont {R\"uegg}, \citenamefont {Coh},\ and\ \citenamefont {Moore}}]{PhysRevB.88.155127}%
  \BibitemOpen
  \bibfield  {author} {\bibinfo {author} {\bibfnamefont {A.}~\bibnamefont {R\"uegg}}, \bibinfo {author} {\bibfnamefont {S.}~\bibnamefont {Coh}}, \ and\ \bibinfo {author} {\bibfnamefont {J.~E.}\ \bibnamefont {Moore}},\ }\href {\doibase 10.1103/PhysRevB.88.155127} {\bibfield  {journal} {\bibinfo  {journal} {Phys. Rev. B}\ }\textbf {\bibinfo {volume} {88}},\ \bibinfo {pages} {155127} (\bibinfo {year} {2013})}\BibitemShut {NoStop}%
\bibitem [{\citenamefont {Ran}\ \emph {et~al.}(2009)\citenamefont {Ran}, \citenamefont {Zhang},\ and\ \citenamefont {Vishwanath}}]{Ran2009}%
  \BibitemOpen
  \bibfield  {author} {\bibinfo {author} {\bibfnamefont {Y.}~\bibnamefont {Ran}}, \bibinfo {author} {\bibfnamefont {Y.}~\bibnamefont {Zhang}}, \ and\ \bibinfo {author} {\bibfnamefont {A.}~\bibnamefont {Vishwanath}},\ }\href {https://doi.org/10.1038/nphys1220} {\bibfield  {journal} {\bibinfo  {journal} {Nature Physics}\ }\textbf {\bibinfo {volume} {5}},\ \bibinfo {pages} {298} (\bibinfo {year} {2009})}\BibitemShut {NoStop}%
\bibitem [{\citenamefont {Wang}\ \emph {et~al.}(2020)\citenamefont {Wang}, \citenamefont {Xue}, \citenamefont {Zhang},\ and\ \citenamefont {Chong}}]{PhysRevLett.124.243602}%
  \BibitemOpen
  \bibfield  {author} {\bibinfo {author} {\bibfnamefont {Q.}~\bibnamefont {Wang}}, \bibinfo {author} {\bibfnamefont {H.}~\bibnamefont {Xue}}, \bibinfo {author} {\bibfnamefont {B.}~\bibnamefont {Zhang}}, \ and\ \bibinfo {author} {\bibfnamefont {Y.~D.}\ \bibnamefont {Chong}},\ }\href {\doibase 10.1103/PhysRevLett.124.243602} {\bibfield  {journal} {\bibinfo  {journal} {Phys. Rev. Lett.}\ }\textbf {\bibinfo {volume} {124}},\ \bibinfo {pages} {243602} (\bibinfo {year} {2020})}\BibitemShut {NoStop}%
\bibitem [{\citenamefont {Ferreira}\ and\ \citenamefont {Mucciolo}(2015)}]{PhysRevLett.115.106601}%
  \BibitemOpen
  \bibfield  {author} {\bibinfo {author} {\bibfnamefont {A.}~\bibnamefont {Ferreira}}\ and\ \bibinfo {author} {\bibfnamefont {E.~R.}\ \bibnamefont {Mucciolo}},\ }\href {\doibase 10.1103/PhysRevLett.115.106601} {\bibfield  {journal} {\bibinfo  {journal} {Phys. Rev. Lett.}\ }\textbf {\bibinfo {volume} {115}},\ \bibinfo {pages} {106601} (\bibinfo {year} {2015})}\BibitemShut {NoStop}%
\bibitem [{\citenamefont {He}\ \emph {et~al.}(2013)\citenamefont {He}, \citenamefont {Zhu}, \citenamefont {Wu}, \citenamefont {Liu}, \citenamefont {Liang},\ and\ \citenamefont {Kou}}]{PhysRevB.87.075126}%
  \BibitemOpen
  \bibfield  {author} {\bibinfo {author} {\bibfnamefont {J.}~\bibnamefont {He}}, \bibinfo {author} {\bibfnamefont {Y.-X.}\ \bibnamefont {Zhu}}, \bibinfo {author} {\bibfnamefont {Y.-J.}\ \bibnamefont {Wu}}, \bibinfo {author} {\bibfnamefont {L.-F.}\ \bibnamefont {Liu}}, \bibinfo {author} {\bibfnamefont {Y.}~\bibnamefont {Liang}}, \ and\ \bibinfo {author} {\bibfnamefont {S.-P.}\ \bibnamefont {Kou}},\ }\href {\doibase 10.1103/PhysRevB.87.075126} {\bibfield  {journal} {\bibinfo  {journal} {Phys. Rev. B}\ }\textbf {\bibinfo {volume} {87}},\ \bibinfo {pages} {075126} (\bibinfo {year} {2013})}\BibitemShut {NoStop}%
\bibitem [{\citenamefont {Black-Schaffer}\ and\ \citenamefont {Balatsky}(2012)}]{PhysRevB.86.115433}%
  \BibitemOpen
  \bibfield  {author} {\bibinfo {author} {\bibfnamefont {A.~M.}\ \bibnamefont {Black-Schaffer}}\ and\ \bibinfo {author} {\bibfnamefont {A.~V.}\ \bibnamefont {Balatsky}},\ }\href {\doibase 10.1103/PhysRevB.86.115433} {\bibfield  {journal} {\bibinfo  {journal} {Phys. Rev. B}\ }\textbf {\bibinfo {volume} {86}},\ \bibinfo {pages} {115433} (\bibinfo {year} {2012})}\BibitemShut {NoStop}%
\bibitem [{\citenamefont {Gonz\'alez}\ and\ \citenamefont {Fern\'andez-Rossier}(2012)}]{PhysRevB.86.115327}%
  \BibitemOpen
  \bibfield  {author} {\bibinfo {author} {\bibfnamefont {J.~W.}\ \bibnamefont {Gonz\'alez}}\ and\ \bibinfo {author} {\bibfnamefont {J.}~\bibnamefont {Fern\'andez-Rossier}},\ }\href {\doibase 10.1103/PhysRevB.86.115327} {\bibfield  {journal} {\bibinfo  {journal} {Phys. Rev. B}\ }\textbf {\bibinfo {volume} {86}},\ \bibinfo {pages} {115327} (\bibinfo {year} {2012})}\BibitemShut {NoStop}%
\bibitem [{\citenamefont {L\"u}\ \emph {et~al.}(2013)\citenamefont {L\"u}, \citenamefont {Lu}, \citenamefont {Shen},\ and\ \citenamefont {Ng}}]{PhysRevB.87.195122}%
  \BibitemOpen
  \bibfield  {author} {\bibinfo {author} {\bibfnamefont {H.-F.}\ \bibnamefont {L\"u}}, \bibinfo {author} {\bibfnamefont {H.-Z.}\ \bibnamefont {Lu}}, \bibinfo {author} {\bibfnamefont {S.-Q.}\ \bibnamefont {Shen}}, \ and\ \bibinfo {author} {\bibfnamefont {T.-K.}\ \bibnamefont {Ng}},\ }\href {\doibase 10.1103/PhysRevB.87.195122} {\bibfield  {journal} {\bibinfo  {journal} {Phys. Rev. B}\ }\textbf {\bibinfo {volume} {87}},\ \bibinfo {pages} {195122} (\bibinfo {year} {2013})}\BibitemShut {NoStop}%
\bibitem [{\citenamefont {Shen}(2017)}]{Shen2017}%
  \BibitemOpen
  \bibfield  {author} {\bibinfo {author} {\bibfnamefont {S.-Q.}\ \bibnamefont {Shen}},\ }\enquote {\bibinfo {title} {Impurities and defects in topological insulators},}\ in\ \href {\doibase 10.1007/978-981-10-4606-3_8} {\emph {\bibinfo {booktitle} {Topological Insulators: Dirac Equation in Condensed Matter}}}\ (\bibinfo  {publisher} {Springer Singapore},\ \bibinfo {address} {Singapore},\ \bibinfo {year} {2017})\ pp.\ \bibinfo {pages} {153--171}\BibitemShut {NoStop}%
\bibitem [{\citenamefont {Juli\`a-Farr\'e}\ \emph {et~al.}(2020)\citenamefont {Juli\`a-Farr\'e}, \citenamefont {M\"uller}, \citenamefont {Lewenstein},\ and\ \citenamefont {Dauphin}}]{PhysRevLett.125.240601}%
  \BibitemOpen
  \bibfield  {author} {\bibinfo {author} {\bibfnamefont {S.}~\bibnamefont {Juli\`a-Farr\'e}}, \bibinfo {author} {\bibfnamefont {M.}~\bibnamefont {M\"uller}}, \bibinfo {author} {\bibfnamefont {M.}~\bibnamefont {Lewenstein}}, \ and\ \bibinfo {author} {\bibfnamefont {A.}~\bibnamefont {Dauphin}},\ }\href {\doibase 10.1103/PhysRevLett.125.240601} {\bibfield  {journal} {\bibinfo  {journal} {Phys. Rev. Lett.}\ }\textbf {\bibinfo {volume} {125}},\ \bibinfo {pages} {240601} (\bibinfo {year} {2020})}\BibitemShut {NoStop}%
\bibitem [{\citenamefont {Mao}\ \emph {et~al.}(2016)\citenamefont {Mao}, \citenamefont {Jiang}, \citenamefont {Moldovan}, \citenamefont {Li}, \citenamefont {Watanabe}, \citenamefont {Taniguchi}, \citenamefont {Masir}, \citenamefont {Peeters},\ and\ \citenamefont {Andrei}}]{Mao2016}%
  \BibitemOpen
  \bibfield  {author} {\bibinfo {author} {\bibfnamefont {J.}~\bibnamefont {Mao}}, \bibinfo {author} {\bibfnamefont {Y.}~\bibnamefont {Jiang}}, \bibinfo {author} {\bibfnamefont {D.}~\bibnamefont {Moldovan}}, \bibinfo {author} {\bibfnamefont {G.}~\bibnamefont {Li}}, \bibinfo {author} {\bibfnamefont {K.}~\bibnamefont {Watanabe}}, \bibinfo {author} {\bibfnamefont {T.}~\bibnamefont {Taniguchi}}, \bibinfo {author} {\bibfnamefont {M.~R.}\ \bibnamefont {Masir}}, \bibinfo {author} {\bibfnamefont {F.~M.}\ \bibnamefont {Peeters}}, \ and\ \bibinfo {author} {\bibfnamefont {E.~Y.}\ \bibnamefont {Andrei}},\ }\href {\doibase 10.1038/nphys3665} {\bibfield  {journal} {\bibinfo  {journal} {Nature Physics}\ }\textbf {\bibinfo {volume} {12}},\ \bibinfo {pages} {545} (\bibinfo {year} {2016})}\BibitemShut {NoStop}%
\bibitem [{\citenamefont {Ugeda}\ \emph {et~al.}(2011)\citenamefont {Ugeda}, \citenamefont {Fern\'andez-Torre}, \citenamefont {Brihuega}, \citenamefont {Pou}, \citenamefont {Mart\'{\i}nez-Galera}, \citenamefont {P\'erez},\ and\ \citenamefont {G\'omez-Rodr\'{\i}guez}}]{PhysRevLett.107.116803}%
  \BibitemOpen
  \bibfield  {author} {\bibinfo {author} {\bibfnamefont {M.~M.}\ \bibnamefont {Ugeda}}, \bibinfo {author} {\bibfnamefont {D.}~\bibnamefont {Fern\'andez-Torre}}, \bibinfo {author} {\bibfnamefont {I.}~\bibnamefont {Brihuega}}, \bibinfo {author} {\bibfnamefont {P.}~\bibnamefont {Pou}}, \bibinfo {author} {\bibfnamefont {A.~J.}\ \bibnamefont {Mart\'{\i}nez-Galera}}, \bibinfo {author} {\bibfnamefont {R.}~\bibnamefont {P\'erez}}, \ and\ \bibinfo {author} {\bibfnamefont {J.~M.}\ \bibnamefont {G\'omez-Rodr\'{\i}guez}},\ }\href {\doibase 10.1103/PhysRevLett.107.116803} {\bibfield  {journal} {\bibinfo  {journal} {Phys. Rev. Lett.}\ }\textbf {\bibinfo {volume} {107}},\ \bibinfo {pages} {116803} (\bibinfo {year} {2011})}\BibitemShut {NoStop}%
\bibitem [{\citenamefont {Chu}\ \emph {et~al.}(2012)\citenamefont {Chu}, \citenamefont {Lu},\ and\ \citenamefont {Shen}}]{Chu_2012}%
  \BibitemOpen
  \bibfield  {author} {\bibinfo {author} {\bibfnamefont {R.-L.}\ \bibnamefont {Chu}}, \bibinfo {author} {\bibfnamefont {J.}~\bibnamefont {Lu}}, \ and\ \bibinfo {author} {\bibfnamefont {S.-Q.}\ \bibnamefont {Shen}},\ }\href {\doibase 10.1209/0295-5075/100/17013} {\bibfield  {journal} {\bibinfo  {journal} {Europhysics Letters}\ }\textbf {\bibinfo {volume} {100}},\ \bibinfo {pages} {17013} (\bibinfo {year} {2012})}\BibitemShut {NoStop}%
\bibitem [{\citenamefont {Santos~Pires}\ \emph {et~al.}(2022)\citenamefont {Santos~Pires}, \citenamefont {Jo\~ao}, \citenamefont {Ferreira}, \citenamefont {Amorim},\ and\ \citenamefont {Viana Parente~Lopes}}]{PhysRevLett.129.196601}%
  \BibitemOpen
  \bibfield  {author} {\bibinfo {author} {\bibfnamefont {J.~P.}\ \bibnamefont {Santos~Pires}}, \bibinfo {author} {\bibfnamefont {S.~M.}\ \bibnamefont {Jo\~ao}}, \bibinfo {author} {\bibfnamefont {A.}~\bibnamefont {Ferreira}}, \bibinfo {author} {\bibfnamefont {B.}~\bibnamefont {Amorim}}, \ and\ \bibinfo {author} {\bibfnamefont {J.~M.}\ \bibnamefont {Viana Parente~Lopes}},\ }\href {\doibase 10.1103/PhysRevLett.129.196601} {\bibfield  {journal} {\bibinfo  {journal} {Phys. Rev. Lett.}\ }\textbf {\bibinfo {volume} {129}},\ \bibinfo {pages} {196601} (\bibinfo {year} {2022})}\BibitemShut {NoStop}%
\bibitem [{\citenamefont {Ni}\ \emph {et~al.}(2020)\citenamefont {Ni}, \citenamefont {Huang},\ and\ \citenamefont {Liu}}]{PhysRevB.101.125114}%
  \BibitemOpen
  \bibfield  {author} {\bibinfo {author} {\bibfnamefont {X.}~\bibnamefont {Ni}}, \bibinfo {author} {\bibfnamefont {H.}~\bibnamefont {Huang}}, \ and\ \bibinfo {author} {\bibfnamefont {F.}~\bibnamefont {Liu}},\ }\href {\doibase 10.1103/PhysRevB.101.125114} {\bibfield  {journal} {\bibinfo  {journal} {Phys. Rev. B}\ }\textbf {\bibinfo {volume} {101}},\ \bibinfo {pages} {125114} (\bibinfo {year} {2020})}\BibitemShut {NoStop}%
\bibitem [{\citenamefont {Tu}\ \emph {et~al.}(2023)\citenamefont {Tu}, \citenamefont {Wu}, \citenamefont {Liu},\ and\ \citenamefont {Li}}]{Tu_2023}%
  \BibitemOpen
  \bibfield  {author} {\bibinfo {author} {\bibfnamefont {W.}~\bibnamefont {Tu}}, \bibinfo {author} {\bibfnamefont {Y.-J.}\ \bibnamefont {Wu}}, \bibinfo {author} {\bibfnamefont {C.-C.}\ \bibnamefont {Liu}}, \ and\ \bibinfo {author} {\bibfnamefont {N.}~\bibnamefont {Li}},\ }\href {\doibase 10.1209/0295-5075/acc41d} {\bibfield  {journal} {\bibinfo  {journal} {Europhysics Letters}\ }\textbf {\bibinfo {volume} {142}},\ \bibinfo {pages} {16002} (\bibinfo {year} {2023})}\BibitemShut {NoStop}%
\bibitem [{\citenamefont {Mook}\ \emph {et~al.}(2014)\citenamefont {Mook}, \citenamefont {Henk},\ and\ \citenamefont {Mertig}}]{PhysRevB.90.024412}%
  \BibitemOpen
  \bibfield  {author} {\bibinfo {author} {\bibfnamefont {A.}~\bibnamefont {Mook}}, \bibinfo {author} {\bibfnamefont {J.}~\bibnamefont {Henk}}, \ and\ \bibinfo {author} {\bibfnamefont {I.}~\bibnamefont {Mertig}},\ }\href {\doibase 10.1103/PhysRevB.90.024412} {\bibfield  {journal} {\bibinfo  {journal} {Phys. Rev. B}\ }\textbf {\bibinfo {volume} {90}},\ \bibinfo {pages} {024412} (\bibinfo {year} {2014})}\BibitemShut {NoStop}%
\bibitem [{\citenamefont {Hatsugai}(1993)}]{PhysRevLett.71.3697}%
  \BibitemOpen
  \bibfield  {author} {\bibinfo {author} {\bibfnamefont {Y.}~\bibnamefont {Hatsugai}},\ }\href {\doibase 10.1103/PhysRevLett.71.3697} {\bibfield  {journal} {\bibinfo  {journal} {Phys. Rev. Lett.}\ }\textbf {\bibinfo {volume} {71}},\ \bibinfo {pages} {3697} (\bibinfo {year} {1993})}\BibitemShut {NoStop}%
\bibitem [{\citenamefont {Haldane}()}]{lec}%
  \BibitemOpen
  \bibfield  {author} {\bibinfo {author} {\bibfnamefont {D.}~\bibnamefont {Haldane}},\ }\href {https://topocondmat.org/w4_haldane/haldane_model.html} {\enquote {\bibinfo {title} {Haldane model, berry curvature, and chern number},}\ }\BibitemShut {NoStop}%
\bibitem [{\citenamefont {Vicarelli}\ \emph {et~al.}(2015)\citenamefont {Vicarelli}, \citenamefont {Heerema}, \citenamefont {Dekker},\ and\ \citenamefont {Zandbergen}}]{Vicarelli2015}%
  \BibitemOpen
  \bibfield  {author} {\bibinfo {author} {\bibfnamefont {L.}~\bibnamefont {Vicarelli}}, \bibinfo {author} {\bibfnamefont {S.~J.}\ \bibnamefont {Heerema}}, \bibinfo {author} {\bibfnamefont {C.}~\bibnamefont {Dekker}}, \ and\ \bibinfo {author} {\bibfnamefont {H.~W.}\ \bibnamefont {Zandbergen}},\ }\href {https://doi.org/10.1021/acsnano.5b01762} {\bibfield  {journal} {\bibinfo  {journal} {ACS Nano}\ }\textbf {\bibinfo {volume} {9}},\ \bibinfo {pages} {3428} (\bibinfo {year} {2015})}\BibitemShut {NoStop}%
\bibitem [{\citenamefont {Robertson}\ \emph {et~al.}(2012)\citenamefont {Robertson}, \citenamefont {Allen}, \citenamefont {Wu}, \citenamefont {He}, \citenamefont {Olivier}, \citenamefont {Neethling}, \citenamefont {Kirkland},\ and\ \citenamefont {Warner}}]{Robertson2012}%
  \BibitemOpen
  \bibfield  {author} {\bibinfo {author} {\bibfnamefont {A.~W.}\ \bibnamefont {Robertson}}, \bibinfo {author} {\bibfnamefont {C.~S.}\ \bibnamefont {Allen}}, \bibinfo {author} {\bibfnamefont {Y.~A.}\ \bibnamefont {Wu}}, \bibinfo {author} {\bibfnamefont {K.}~\bibnamefont {He}}, \bibinfo {author} {\bibfnamefont {J.}~\bibnamefont {Olivier}}, \bibinfo {author} {\bibfnamefont {J.}~\bibnamefont {Neethling}}, \bibinfo {author} {\bibfnamefont {A.~I.}\ \bibnamefont {Kirkland}}, \ and\ \bibinfo {author} {\bibfnamefont {J.~H.}\ \bibnamefont {Warner}},\ }\href {https://doi.org/10.1038/ncomms2141} {\bibfield  {journal} {\bibinfo  {journal} {Nature Communications}\ }\textbf {\bibinfo {volume} {3}},\ \bibinfo {pages} {1144} (\bibinfo {year} {2012})}\BibitemShut {NoStop}%
\bibitem [{\citenamefont {Jiang}\ \emph {et~al.}(2019)\citenamefont {Jiang}, \citenamefont {Xu}, \citenamefont {Lu}, \citenamefont {Sun},\ and\ \citenamefont {Ni}}]{doi:10.34133/2019/4641739}%
  \BibitemOpen
  \bibfield  {author} {\bibinfo {author} {\bibfnamefont {J.}~\bibnamefont {Jiang}}, \bibinfo {author} {\bibfnamefont {T.}~\bibnamefont {Xu}}, \bibinfo {author} {\bibfnamefont {J.}~\bibnamefont {Lu}}, \bibinfo {author} {\bibfnamefont {L.}~\bibnamefont {Sun}}, \ and\ \bibinfo {author} {\bibfnamefont {Z.}~\bibnamefont {Ni}},\ }\href {\doibase 10.34133/2019/4641739} {\bibfield  {journal} {\bibinfo  {journal} {Research}\ }\textbf {\bibinfo {volume} {2019}} (\bibinfo {year} {2019}),\ 10.34133/2019/4641739}\BibitemShut {NoStop}%
\end{thebibliography}
\end{document}